\documentclass[aps, preprintnumbers, tightenlines, nofootinbib, 11pt]{revtex4}

\usepackage{amssymb,amsmath,amsfonts,amsbsy,epsfig,wrapfig,caption,cancel,graphicx, pstricks, slashed}

\newcommand{\eq}[2]{\begin{equation}\label{#1}#2 \end{equation}}
\newcommand{\ft}[1]{ \frac{\dot #1}{#1}}

\newcommand{\fp}[1]{ \frac{#1'}{#1}}
\newcommand{\fpp}[1]{ \frac{#1''}{#1}}

\usepackage{xcolor}

\newcommand{\p}{\varphi}
\newcommand{\w}{\widetilde}
\newcommand{\mpl}{M_{(5)}}
\newcommand{\mplf}{M_{\rm pl}}

\newcommand{\calL}{{\cal L}}
\newcommand{\calP}{{\cal P}}
\newcommand{\calR}{{\cal R}}

\newcommand{\calA}{{\cal A}}

\def\bea{\begin{eqnarray}}
\def\eea{\end{eqnarray}}
\def\be{\begin{equation}}
\def\ee{\end{equation}}
\def\ba{\begin{array}}
\def\ea{\end{array}}

\def\nn{\nonumber}

\begin{document}
\author{Ignatios Antoniadis$^{1,2}$}
\email{ignatios.antoniadis@upmc.fr}
\author{Subodh P. Patil$^{3}$}
\email{subodh.patil@unige.ch}
\affiliation{1) LPTHE Jussieu, UMR CNRS 7589, Sorbonne Universit\'es, UPMC Paris 6,\\ 75005 Paris, France}
\affiliation{2) Albert Einstein Center for Fundamental Physics, ITP University of Bern, Sidlerstrasse 5\\ CH-3012 Bern, Switzerland\\}
\affiliation{3) Dept. of Theoretical Physics, University of Geneva, 24 Quai Ansermet, CH-1211 Geneva-4, Switzerland\\}
\date{\today}

\title{The Effective Strength of Gravity, the Scale of Inflation\\ (and how KK gravitons evade the Higuchi Bound)}

\begin{abstract}
For any given momentum transfer, gravitational interactions have a strength set by a characteristic scale $M_*$ inferred from amplitudes calculated in an effective theory with a strong coupling scale $M_{**}$. These are in general different from each other and $\mplf$, the macroscopic strength of gravity as determined by (laboratory scale) Cavendish experiments. During single field inflation, $M_*$ can differ from $\mplf$ due to the presence of any number of (hidden) universally coupled species between laboratory and inflationary scales. Although this has no effect on dimensionless (i.e. observable) quantities measured at a fixed scale such as the amplitude and spectral properties of the CMB anisotropies, it complicates the \textit{inference} of an absolute scale of inflation given any detection of primordial tensors. In this note we review and elaborate upon these facts and address concerns raised in a recent paper. 
\end{abstract}

\maketitle
\vspace{-20pt}
\section{Introductory Remarks}

Below the scale at which strong gravitational effects become relevant, gravity can be treated as an effective theory \cite{Donoghue, Burgess}. In pure Einstein gravity, this strong coupling scale is simply given by the reduced Planck mass $\mplf = 2.435\times 10^{18}$ GeV. In the presence of matter, the strong coupling scale is lowered as
\eq{scs}{M_{**} \sim \mplf/\sqrt{N}}
where $N$ is shorthand for a weighted index that counts the total numbers of particles of different spins present.  As reviewed in \cite{Antoniadis:2014xva}, one can understand this result in a variety of ways\footnote{See \cite{Dvali3, Dvali1, Dvali2} for a range of arguments including black hole thermodynamics.} for which we offer our own simple derivation in the next section. On the other hand, the effective strength of gravity -- $M_*$ -- is in principle an independent quantity. One determines $M_*$ via local scattering experiments, say a test (point) mass scattering off of a heavier mass. As we also review shortly, depending on the process in question  and its scale\footnote{i.e. Allowing for interactions that violate the equivalence principle at high energies.},
\eq{}{M_* \sim \mplf/\sqrt{N_*}} 
counts the number $N_*$ of species that can mediate \textit{tree-level} interactions of gravitational strength. During inflation for example in the presence of large (but stabilized) extra dimensions, processes that couple only to the transverse traceless (TT) polarizations of the graviton will turn up the scale corresponding to $N_* = 1 + N^{\rm T}_{\rm KK}$ where $N^{\rm T}_{\rm KK}$ counts the number of TT Kaluza-Klein (KK) resonances with masses below the momentum transfer in question. Processes that couple only to the longitudinal polarization of the graviton\footnote{Recalling that in unitary gauge, spacetime is foliated such that inflaton fluctuations are gauged away. The graviton thus acquires a propagating longitudinal polarization by `eating' the fluctuating inflaton \cite{Cheung:2007st} (reviewed in \cite{Chluba}).} (equivalently, processes mediated by species that couple to the trace of the energy momentum (EM) tensor) on the other hand see the scale corresponding to $N_*$ where $N_* = 1 + \w N_*$ where $\w N_*$ effectively counts the (non KK) \textit{universally coupled} species contributing to the process in question in addition to any KK graviscalars whose couplings or expectation values may have shifted between laboratory scales and the scale of inflation. \textit{In what follows, the term universally coupled specifically refers to all species that can mediate tree-level exchange processes between covariantly conserved sources.}

In this note, we further elaborate upon these facts and their consequences for \textit{inferring} absolute scales from observable quantities during inflation. In doing so, we address various concerns raised in a recent paper \cite{Kleban}, where firstly it was reasoned that KK gravitons could not contribute towards lowering $M_*$ during inflation since this would require their masses to be less than the Hubble scale, seemingly in violation of the Higuchi bound \cite{Higuchi}. A straightforward corollary of this reasoning however, would be that it is impossible for the compactification scale to be less than the effective cosmological constant, implying a powerful no go theorem for compactifications were it true. We demonstrate by explicit example and proof that as expected from the underlying consistency of higher dimensional Einstein gravity, that KK gravitons evade the Higuchi bound \footnote{Compactifications spontaneously break higher dimensional Lorentz invariance. As a result, the underlying healthiness of the higher dimensional theory manifests in cancellations that preclude the problematic terms that would otherwise have implied a propagating ghost in a given mass range on a de Sitter (dS) background (see appendix).}.

Secondly, it was reasoned in \cite{Kleban} that universally coupled species that are not TT spin-2 excitations could not affect the inferred scale of inflation from a positive detection of primordial tensors, since the former couldn't have generated perturbations that linearly source B-mode polarization in the cosmic microwave background (CMB) \cite{polnarev}. This reasoning is inaccurate since it neglects to treat gravity as an effective theory and (as we remind the reader shortly) results in the inconsistent corollary that it would be possible to infer a scale for inflation beyond the scale at which a classical description of spacetime breaks down. A more careful examination of how one connects theoretical quantities to what is extracted from the CMB confirms the basic conclusions of \cite{Antoniadis:2014xva}; that inferring an absolute scale of inflation is complicated by the uncertainty in $M_*$:
\eq{}{V_*^{1/4} \sim \frac{r_*^{1/4}}{\sqrt N_*}\,3.28 \times 10^{16}\,\rm{GeV}}
where $r_*$ is the observed tensor to scalar ratio and $N_* = (1 + N^{\rm T}_{\rm KK})(1 + \w N_*)$ with the caveat that one can only trust this inference if the resulting curvature is such that $R \lesssim \mplf^2/N$, or  
\eq{hbound}{\frac{H^2_*}{\mplf^2} \lesssim \frac{1}{N}}
with $N$ being the total number of species in our theory and $H_*$ is the Hubble factor during inflation (the above is nothing more than the obvious requirement that $H_*^2/M_{**}^2 \lesssim 1$) \cite{Antoniadis:2014xva}. Furthermore, as we elaborate upon further, non KK universally coupled species can violate the so-called single field tensor to scalar consistency relation even as the spectral properties of the anisotropies remain unchanged. 

\subsection{Outline}

Some of our findings are direct corollaries of results established elsewhere \cite{Dvali3,Dvali2,Dvali1} while others are obvious in hindsight although evidently obscured by a number of moving parts. Therefore, we err on the side of detail in the following treatment, clarifying various details omitted in \cite{Antoniadis:2014xva} and presenting the observation on the nature of the Higuchi bound alluded to in the title. Readers interested only in the latter result and any wider lessons to be drawn are invited to skip to the conclusion and appendices. We begin by reviewing aspects of how short distance gravity can differ from macroscopic gravity in a general context after which we fix to the specific setting of inflationary cosmology. We derive the implications of $M_*$ and $M_{**}$ differing from the macroscopic strength of gravity (set by $\mplf = 2.44 \times 10^{18}$ GeV) for cosmological observables and in particular, inferring an absolute energy scale for inflation, after which we address concerns raised in \cite{Kleban} and conclude. Along the way, we draw attention to the fact that the process dependence of $M_*$ can result in a violation of the single-field tensor to scalar consistency relation, in addition to the fact that any positive detection of primordial tensors necessarily bounds the \textit{total} number of species in the universe, hidden or otherwise. 

\section{Review}
\subsection{The different scales of gravity}
In the absence of matter, gravity becomes strongly coupled as the momentum transfer for any given process approaches the (reduced) Planck scale $\mplf = 2.435\times 10^{18}$ GeV or when the background curvature approaches $R \sim \mplf^2$. In the presence of matter, the strong coupling scale is lowered as $M_{**} = \mplf/\sqrt{N}$, where $N$ is shorthand for a weighted sum that counts the total number of species present. A direct understanding is arrived at, for example, from the effective action one obtains after integrating out an arbitrary particle spectrum of massive particles minimally coupled to gravity, initially described by the Einstein-Hilbert action:
\eq{S1}{S = \int d^4x\sqrt{-g}\left(\frac{M^2_B}{2}R - \Lambda_B + \calL_M\right) }
where $\calL_M$ is some arbitrary matter sector and $M_B$ and $\Lambda_B$ denote recognizable dimensionful couplings that are subject to renormalization. All graviton scattering amplitudes (or on shell sub-amplitudes) can be calculated from the effective action that results from integrating out the matter fields. To one loop, these amplitudes are reproduced by the effective action
\eq{1l}{ S = \int d^4x\sqrt{-g}\left(\frac{\mplf^2}{2}R - \Lambda + c_1 R^2 + c_2 R_{\mu\nu}R^{\mu\nu}\right) + ... }
where the ellipses denote higher order contributions in the curvature and loop expansion. The coefficient of the Einstein-Hilbert term and the cosmological constant are fixed by measurements with the `bare' terms in (\ref{S1}) absorbing the divergences after dimensionally regularizing, and where 
\eq{}{c_{1,2} \sim \sum_i \frac{w_i}{16\pi^2}\, {\rm log}\frac{m_i^2}{\mu^2}}
represents a weighted sum over all massive species present with the $w_i$ calculable in terms of the spins of the particles integrated out \cite{Birrell, Vassilevich:2003xt}. Given that $c_{1,2} \sim N$ we see from counting derivatives that the perturbative expansion breaks down when the momentum transfer approaches
\eq{}{p^2 \sim \mplf^2/N}
or when the background curvature approaches
\eq{mc}{R \sim \mplf^2/N}
both of which are manifestations of the fact in the presence of an arbitrary spectrum of particles, the scale at which strong gravity effects become relevant is set by (\ref{scs}) -- $M_{**} \sim \mplf/\sqrt N$. Crucially, we note that \textit{the maximum allowed curvature before a classical description of spacetime breaks} down is given by $\mplf^2/N$ instead of $\mplf^2$ as would be the case in a purely gravitational theory, a fact that is not without consequence for inflationary cosmology.

On the other hand, the effective strength of gravity at any given scale, set by $M_*$, is an independent quantity. Although all massive species lower the strong coupling scale as (\ref{scs}), only those which mediate tree level interactions between covariantly conserved sources can enhance the strength of gravitational interactions (the latter being our definition of \textit{universally coupled} species) immediately below distance scales smaller than their inverse Compton wavelength\footnote{This is because the loop threshold effects that lower the strong coupling scale only become relevant when all higher loop corrections also become relevant, i.e. when the momentum transfer approaches $M_{**}$, precisely when the effective theory starts to break down \cite{Antoniadis:2014xva}.}. This occurs independently of the process for Kaluza-Klein (KK) resonances and in a process dependent (i.e. equivalence principle violating) manner for species that couple to the trace of the EM tensor of the source:
\eq{sg}{M_* \sim \mplf/\sqrt{N_*}}
where $N_*$ counts the number of species contributing to (and with masses below) the momentum transfer of the tree level process in question. As an example of the latter, one can contemplate (KK) graviscalars, 4-dimensional (4D) scalars or vector bosons with explicit non-minimal couplings to gravity (hence coupling to the trace of the energy momentum tensor of any source) or which mediate effective interactions via higher dimensional operators -- e.g. dimension 5 in the case of pseudo-scalar exchange \cite{Adelberger:2003zx} or dimension 6 in the context of Higgs effective field theory \cite{Antoniadis:2014xva}\footnote{One could also contemplate that $M_* \equiv \mplf$ for all momentum transfer up to the scale $M_{**}$ as was done in \cite{Gasperini}, where $M_{**}$ can be lowered arbitrarily by engineering an appropriate field content.}. 

One can understand the lowering of $M_*$ immediately above the threshold $M$ where the latter denotes the mass of the species by first considering what happens for a KK graviton. For a given TT KK resonance with mass $M$, that tree level graviton exchange between any two conserved sources is augmented as (suppressing tensor structure for simplicity):
\begin{figure}[h!]\epsfig{file=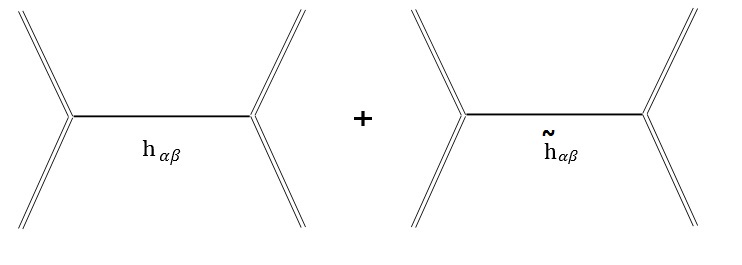, height=1.2in, width=3.3in}\end{figure} 
\eq{tl}{\frac{1}{\mplf^2 p^2} \to \frac{1}{\mplf^2 p^2} + \frac{g}{\mplf^2 (p^2+M^2)}}
where $g$ counts the number of contributing KK polarizations of mass $M$ (as illustrated above). Hence in the regime $M^2 \ll p^2 \ll \mplf^2/N$ we remain within the regime of validity of our effective theory and the tree level exchange effectively becomes
\eq{mple}{\frac{1}{\mplf^2 p^2} \to \frac{g+1}{\mplf^2p^2}}
implying an increase in the strength of gravity at distances smaller than $M^{-1}$ but greater than $M_{**}^{-1}$ as per (\ref{sg}). Consider now the effects of any other universally coupled (non KK) species, for example the Higgs via the dimension 6 effective operators coupled to scalar or fermionic matter:
\eq{ucs}{\Delta\calL_{\rm eff} \sim c_1\frac{H^\dag H}{\mplf^2}\partial_\mu\p\partial^\mu\p + c_2\frac{H^\dag H}{\mplf^2}\bar\psi\slashed\partial\psi\sim  c_{\{1,2\}}\frac{H^\dag H}{\mplf^2}T^\mu_\mu}
which evidently couples the singlet component of the Higgs $h$ to the trace of the energy-momentum tensor of any conserved source made up of $\phi$ or $\psi$ quanta with the effective interaction
\eq{d6}{\Delta\calL_{\rm eff}\sim c_i\frac{v\,h}{\mplf^2}T^\mu_\mu}
where $v$ denotes the vacuum expectation value (vev) of the Higgs in the phase we're doing perturbation theory in. One finds an identical enhancement to gravitational interactions as in (\ref{tl}) with the replacement 
\eq{gdef}{g \to c^2_i v^2/\mplf^2.} Similar enhancements can come from KK graviscalars -- when more than one extra dimension is compactified, there are always extra scalar polarizations $H^{(n)}$ distinct from the scalar mode of the massive 4D graviton which couple to the trace of the EM tensor: 
\eq{d5}{\Delta\calL_{\rm eff}\sim \sum_{n=1}\kappa\frac{H^{(n)}}{\mplf}T^\mu_\mu}
with $\kappa$, a constant numerical factor \cite{Giudice}. As soon as the momentum transfer exceeds the mass of any of these scalar KK resonances, one finds an enhancement akin to (\ref{mple}) with $g \to \kappa^2$ per resonance. One thus concludes that for Cavendish experiments performed with point masses, $M_* \sim \mplf/\sqrt{N_*}$ where $N_*$ stands for a process dependent weighted index\footnote{We note in passing that there are certain caveat emptors and ambiguities when dealing with scale dependent quantities in EFT of gravity (see discussions in \cite{Anber,NBB}). We evade these issues by dealing only with (unambiguously defined) physically observable quantities such as on-shell S-matrix elements.}.

\subsection{The scale of Inflation}

So what, if any consequences do the facts reviewed above have for cosmological observations involving curvature and tensor perturbations? Not so many as it turns out, since observable quantities are always dimensionless and therefore independent of the units in which they are expressed (Planck units being the naturally available scale in early universe cosmology). The amplitude and spectral properties of the CMB anisotropies in particular are unaffected by any of the considerations above. However, in trying to \textit{infer} an energy scale of inflation, one necessarily runs up against the fact that $M_* \neq \mplf$, and that the absolute scale of inflation is uncertain up to our lack of knowledge of $M_*$ beyond laboratory scales.

To see this, we first fix a particular context for the purposes of demonstration even though the results generalize \cite{Antoniadis:2014xva}. From the outset, we stress that we work in the context of single field inflation. Any other fields present therefore couple to the inflaton only through gravitational strength interactions. Secondly, we work in a 4D context, even though we allow for KK gravitons to be excited. That is, KK masses are set by the size of extra dimensions (i.e. the expectation value of the moduli that parametrize them) which is an independent parameter from the masses of these moduli, given some stabilization mechanism. Therefore we can consider situations where
\eq{mkkc}{m^2_{\rm KK} < H_*^2 \ll \mu^2}
where $m_{\rm KK}$ is the characteristic KK mass, $\mu$ is characteristic moduli mass that can be taken to be arbitrarily large and $H_*$ is the Hubble factor during inflation. This permits an effectively 4D description of the background over which KK modes with masses up to $H_*^2$ can be excited. Before deriving our desired results, it is useful to have an overview of the different moving parts at work:
\begin{itemize}
\item Cavendish experiments fix $M_*$ to be $M_* = \mplf = 2.44\times 10^{18}$ GeV up to the percent level at scales $\sim 10^{-4}$ m $\sim 10^{-2}$ eV$^{-1}$ \cite{Kapner:2006si}.
\item  For each mass threshold crossed between $10^{-2}$ eV up to the scale of inflation\footnote{Corresponding to a new tree level exchange channel opening up for the process in question.}, the strength of gravity increases as per (\ref{mple}), after which the usual logarithmic running sets in. 
\item The species that contribute depend on the process considered, but always include KK gravitons, consistently treated as massive spin-2 excitations over a 4D solution since $H_*^2 \ll \mu^2$.   
\end{itemize}
In order to illustrate the physics at work in as simple a context as possible, we further make the assumptions that:
\begin{itemize}
\item We either have one extra dimension, so that aside from the zero modes, there are no extra graviscalars/vectors other than those `eaten' by the massive spin-2 KK modes;  
\item or for more than one extra dimension, the mechanism that gives masses to the zero modes of the radion and vector moduli also generates commensurate masses for their KK excitations.
\end{itemize} 
The latter two conditions can readily be relaxed, although for the purposes of simple illustration ensure through (\ref{mkkc}) that the scalar and vector fluctuation modes of the extra dimensions have masses much larger than $H_*$ thus permitting an effective 4D description for the perturbations with no other hidden scalars. For more than one extra dimension, relaxing the latter requirement would result in additional light universally coupled hidden fields which still require a mechanism to generate masses for them to avoid fifth force constraints \footnote{As for non-universally coupled fields, $N$ species of the former alone would fix $M_* \equiv \mplf$ all the way up to $M_{**} = \mplf/\sqrt N$, logarithmic running aside (cf. \cite{Gasperini}). Within this context, see also \cite{Ozsoy:2014sba, Namba:2015gja} where additional fields and interactions typically generate non-Gaussianities.}. 

Since the only dynamical field that has any background time dependence (and energy density) in this set-up is the inflaton, the universe has only one physical clock. Therefore, all perturbations are adiabatic and we are entitled to foliate spacetime such that the inflaton fluctuations are gauged away (comoving/ unitary gauge) \cite{Cheung:2007st}. Since we are only interested in the scalar and tensor perturbations, and since all moduli have masses much greater than any other scale in the problem, the metric induced on the spatial hypersurfaces can be parametrized as  
\eq{mpdec}{h_{ij}(t,x) = a^2(t)e^{2\calR(t,x)}\hat h_{ij};~~~ \hat h_{ij} := \sum_{n}{\rm exp}[\gamma^{(n)}_{ij}],~~~\partial_i \gamma^{(n)}_{ij} = \gamma^{(n)}_{ii} = 0}
where the $\gamma^{(n)}_{ij}$ correspond to the $n^{\rm th}$ induced TT KK resonance which propagates freely at the scale of inflation (i.e. with masses less than the Hubble scale)\footnote{St\"uckelberg decomposing the propagating massive spin-2 graviton as $\hat h^{(n)}_{ij} = \gamma^{(n)}_{ij} + \partial_i A_j + \partial_j A_i + 2\partial_i\partial_j \psi$, the vector perturbation $A_i$ decouples at linear order and the longitudinal polarization $\psi$ has no evolving background and therefore contributes vanishingly to the observed adiabatic mode $\calR$ (\ref{mpdec}).}. If there are $N^{\rm T}_{\rm KK}$ of these resonances, then clearly this would result in a total power for the tensor spectrum of
\eq{tspec}{\calP_\gamma := 2(1 + N^{\rm T}_{\rm KK})\frac{H^2_*}{\pi^2 \mplf^2},}
equivalent to the replacement $M^2_{*T} = \mplf^2/(1 + N^{\rm T}_{\rm KK})$ (where the subscript is to emphasize the process dependence of this quantity). Next, we consider the effect on the curvature perturbations of a scalar $\eta$ that couples to the trace of the energy momentum tensor (equivalently, with non-minimal coupling)\footnote{This incorporates the example of dimension 6 effective interactions (\ref{ucs}) which can be converted into a non-minimal coupling to gravity via field redefinition using the background equations of motion \cite{Weinberg} if the coupling is to the matter component sourcing the background.}:
\eq{jf}{\Delta\calL_{\rm eff}\sim  \xi\,\eta^2\frac{T^\mu_\mu}{\mplf^2} \equiv - \xi\, \eta^2 R}
which is to be considered in conjunction with the original Einstein-Hilbert term
\eq{jf2}{\calL_{\rm eff} \supset \frac{\mplf^2}{2}\left(1 - \frac{2\xi \eta^2}{\mplf^2} \right)R}
One might be tempted to immediately infer from the above that doing perturbation theory around a background where $\eta$ has a non-vanishing expectation value $v$ relative to one where it vanishes, would imply an effective change in the strength of gravity encoded by $M^2_* = \mplf^2/(1 + g)$, with $g :=  2\xi v^2/\mplf^2 \ll 1$ (cf. (\ref{gdef})) by assumption\footnote{A concrete example of this occurs in the context of Higgs Inflation \cite{Bezrukov} where $\eta$ is identified with the inflaton itself (the singlet component of the Higgs), and where the vacuum expectation value (vev) during inflation $v \sim \mplf/\sqrt{-\xi}$ with $\xi \sim - 10^3$ is many orders of magnitude greater than its value in the EW vacuum where $v = 246$ GeV.}. As far as the curvature corrections are concerned, this is essentially correct, although we have to work a little harder to prove this. We first go to the Einstein frame via the conformal transformation 
\eq{ct}{g_{\mu\nu} = \left(1 - \frac{2\xi \eta^2}{\mplf^2} \right)^{-1}{\w g}_{\mu\nu}\, := F^{-1}(\eta){\w g}_{\mu\nu}}
which rescales (\ref{jf2}) into the usual Einstein-Hilbert term 
\eq{ctl}{ \mathcal L_{\rm eff} \supset \frac{\mplf^2}{2}\w R + F^{-2}(\eta)\mathcal L_m\left[F^{-1}(\eta)\w g_{\mu\nu}, \psi,A_\mu,\phi,\eta\right] + ...}
where $\calL_m$ is the Lagrangian that describes all matter content including standard model fields, the inflaton and the species $\eta$. We first observe that all massless fermions and $U(1)$ gauge fields in 4D are conformally invariant, so the transformation (\ref{ct}) has no effect on the conformally rescaled (and correspondingly canonically normalized) fields. Scalar fields on the other hand (unless non-minimally coupled with $\xi = -1/6$) are not. Therefore in the Electroweak (EW) vacuum (where the conformally rescaled Higgs vacuum expectation value dictates all particle masses), Cavendish experiments will turn up $M_* = 2.44 \times 10^{18}$ GeV as the strength of gravity, say at sub-mm scale torsion balance experiments \cite{Kapner:2006si}. The canonically normalized Higgs field is given by $\w H = F^{-1/2}H$, so that all particle masses scale $\propto F^{1/2}(\eta)$ under constant shifts of $\eta$. Equivalently, keeping particle masses fixed, this is equivalent to changing the strength of gravity\footnote{One can also see this from explicitly verifying that the EM tensor derived from (\ref{ctl}) scales as $T^\mu_{\nu}(\eta_*) = \frac{F^2(\eta_0)}{F^2(\eta_*)}T^\mu_\nu(\eta_0)$ which follows directly from its conformal dimension. This is equivalent to keeping the EM tensor fixed, but scaling $M_*$ as $M_* \propto F(\eta)$.} as $M_*(\eta) = F(\eta)/F(\eta_0)\cdot 2.44\times 10^{18}$ GeV where $\xi\eta^2$ in the EW vacuum is defined to be $\xi_0 v_0^2$. If during inflation however, the background shifts to $\xi_*v_*^2$ (either through explicit shifts in the expectation value of $\eta$ or through running of $\xi$), it is straightforward to see that the amplitude of curvature perturbations one infers from (\ref{ctl}) will be given by 
\eq{cspec}{\calP_\calR :=  \frac{H^2_*}{8\pi^2 M_{* s}^2 \epsilon_*};~~~ \epsilon_* := -\dot H_*/H_*^2,}
where $M_{*s}$ above is 
\eq{mstardef}{M_{*s} = F(\eta_*)/F(\eta_0)\cdot  2.44\times 10^{18}\,{\rm GeV}.} 
This amplitude is \textit{fixed} by the observed anisotropies of the CMB to be $\calP(k_*) = \calA \times 10^{-10}$ where $\calA \sim 22.15$ \cite{planckcos}. The tensor to scalar ratio is also a quantity that would be fixed by any putative measurement of primordial tensor modes, and is given replacing $\mplf$ with $M_{*s}$ in (\ref{tspec}) as specified in (\ref{mstardef}) and given (\ref{cspec}) to be
\eq{rfix}{r_*:= \frac{\calP_\gamma}{\calP_\calR} = 16\epsilon_*\left(1 + N^{\rm T}_{\rm KK}\right) = 16\epsilon_*\left(\frac{M^2_{*s}}{M^2_{*T}}\right),}
now with $M^2_{*T} = M^2_{*s}/(1 + N^{\rm T}_{\rm KK})$. Any positive determination of $r_*$ fixes $\epsilon_*$ and implies that in the regime we can trust our effective theory, the scale of inflation is given by
\eq{sfix}{H_*^2 =  M^2_{*T}\left(\frac{\pi^2\mathcal A r_*}{2 \cdot 10^{10}}\right);~~~V_*^{1/4} = M_{*T}\left(\frac{3\pi^2\mathcal A r_*}{2 \cdot 10^{10}}\right)^{1/4};~~~M^2_{*T} = M^2_{*s}/(1 + N^{\rm T}_{\rm KK})} 
with 
\eq{}{M^2_{*s} = \mplf^2\frac{F^2(\eta_*)}{F^2(\eta_0)}\approx \frac{\mplf^2}{1 + \w N_*}}
with $\w N_* := \sum_i g_i$ where for example, $g_i = 2\Delta( \xi_i\eta_i^2)/\mplf^2$ or $g_i = 2\Delta(\xi_i\eta)/\mplf$ for dimension 5 or 6 couplings to the trace of the EM tensor (e.g. (\ref{d6}) and (\ref{d5})) presuming these individual shifts to be small. All of this is subject to the caveat that the inferred Hubble scale is necessarily bounded from above by  (\ref{hbound})
\eq{slim}{H^2_* \lesssim \frac{\mplf^2}{N}}  
where to reiterate, $N$ is the weighted index corresponding to the total number of species present, hidden or otherwise. For example, in the scenario of \cite{Dvali3} with $10^{32}$ hidden copies of the standard model, it would not be possible to infer a scale of inflation greater than a TeV. Furthermore, it is amusing to note that (\ref{sfix}) and (\ref{slim}) together imply a bound on the total number of species in the universe (universally coupled or otherwise) given the fixed amplitude of curvature perturbations in conjunction with any positive determination of $r_*$ that goes as:
\eq{}{N \leq \frac{9.15}{r_*}\times 10^7\frac{\mplf^2}{M^2_{*T}}.}
In the standard scenario, $M^2_{*T} \equiv \mplf^2$ which for any detection of $r_* \sim \mathcal O(10^{-1})$ would imply $N \lesssim 10^9$, and for any given extra dimensional scenario, the factor $\mplf^2/M^2_{*T}$ is a calculable geometrical quantity that encapsulates the size of the extra dimensions \cite{Antoniadis:2014xva}. 

In summary, we see that although the amplitude and spectral properties of CMB anisotropies are unaffected by any difference in the strength of gravity during inflation, \textit{inferring} an absolute scale is complicated by our lack of knowledge of the scale $M_{*T}$ and and $M_{**}$, rendering it effectively uncertain. We further note that since the tilt of the power spectrum of all of the individual graviton polarizations is still fixed by the deviation of the background from an exactly dS geometry $n_T = -2\dot H_*/H^2_*$, we see that (\ref{rfix}) implies a deviation from the tensor to scalar consistency relation:
\eq{csv}{n_T = -\frac{r_*}{8}\left(\frac{M^2_{T*}}{M^2_{s*}}\right).}
which is the only \textit{observable} consequence of the process and scale dependence of the strength of gravity at the scale of inflation.  

\subsection{Response to arXiv:1508.01527}
Recently several concerns regarding some of the results discussed above were raised in \cite{Kleban} which we presently wish to address. Firstly, it was observed that the Higuchi bound nominally appears to forbid the presence of massive spin-2 excitations over a dS background within the mass range
\eq{}{0 \leq m^2_{\rm KK} \leq 2H^2}
which would imply that KK resonances could not be excited during inflation under the circumstances described in the previous subsection and correspondingly affect the inferred scale of inflation. However if such a bound were to truly apply to KK gravitons, one would have a no-go theorem for compactifications that would forbid consistent solutions to higher dimensional Einstein gravity with an effective 4D Hubble scale that is greater than the compactification scale. No such no-go theorem exists\footnote{Were it to do so, higher dimensional Einstein gravity would have to posses at least one other characteristic scale in addition to the higher dimensional Planck mass.}. We demonstrate this by explicit example in appendix A, where we construct a solution to 5D Einstein gravity on an orbifold topology with an empty bulk and bounding branes of opposite tensions that support an induced 4D dS geometry:
\eq{5ddec0}{ds^2 = \frac{\left(1 + H|y|\right)^2}{H^2\tau^2}\left( -d\tau^2 + dx_1^2 + dx_2^2 + dx_3^2\right) + dy^2 } 
where 
\eq{hans0}{H = -\kappa_5^2\Lambda_0/6}
with $\Lambda_{0}, \Lambda_c$ being the tensions of the brane at $y=0$ and $y_c$ respectively with $\Lambda_0$ taken to be negative so that one has an expanding induced dS solution, with the two tensions related by the junction conditions at $y_c$ as $\Lambda_c = -\Lambda_0/(1 + H|y_c|)$. By adjusting $\Lambda_c$, one can thus keep the induced Hubble factor fixed whilst simultaneously dialing the physical inter-brane separation to be as large as one desires
\eq{}{|y_c| = \frac{6(\Lambda_0 + \Lambda_c)}{\kappa_5^2\Lambda_0\Lambda_c}.} Hence in keeping $H^2 = \Lambda_0^2\kappa_5^4/36$ fixed, our solution consistently attains $H^2 \gg \frac{1}{y_c^2}$, and given that the KK mass spectrum scales up to pre-factors of order unity as 
\eq{}{m^2 \sim \pi^2\frac{n^2}{y_c^2}}
we find an explicit construction where the masses of an arbitrary but finite number of KK modes can be made less than the Hubble scale. In appendix B we understand precisely how KK gravitons \textit{evade} Higuchi's bound in this context. The reason for this is straightforward and has to do with the fact that compactifications necessarily spontaneously break higher dimensional Lorentz invariance. As a result, background sources contribute terms that precisely cancel what would have been problematic terms in the equation of motion for the massive spin-2 graviton, as could have been inferred from the outset by the healthiness of higher dimensional Einstein gravity.

Furthermore, it was argued in \cite{Kleban} that lower spin particles could not affect the observed spectrum of tensor perturbations since the latter can only be generated from the Einstein Hilbert term in the effective action:
\eq{ttact}{S = \frac{\mplf^2}{2}\int\sqrt{-g}R.}
This statement is inaccurate, as taking the above as the only source of TT petrurbations leads to the contradiction that it would be possible to infer a scale for inflation beyond where a classical description of geometry breaks down (\ref{mc}). Focussing presently on a strictly 4D context for clarity of discussion, one needs only to realize that in fact it is not (\ref{ttact}), but the effective action\footnote{Where $\phi$ is the inflaton that sources the background evolution.} 
\eq{1li}{ S = \int d^4x\sqrt{-g}\left(\frac{\mplf^2}{2}R -\frac{1}{2}\partial_\mu\phi\partial^\mu\phi - V(\phi) + c_1 R^2 + c_2 R_{\mu\nu}R^{\mu\nu}\right) + ... }
that one has to work with in extracting the tensor power spectrum where as reviewed in the previous section, $c_{1,2} \sim N$ are spin dependent weighted indices that count the \textit{total} number of massive species of all  spins present and where the ellipses denote higher order terms in the curvature and loop expansion (with each independent loop momenta contributing a factor of $N$). On a given background, the resulting quadratic action for the TT polarizations of the graviton is given by
\eq{stt}{S_{TT} = \frac{\mplf^2}{8}\int d^4x\sqrt{-g_{0}}\left[ \dot h_{ij}\dot h_{ij} - \frac{1}{a^2}\partial_k h_{ij}\partial_k h_{ij}\right]\left(1 + c\frac{H_*^2}{\mplf^2} + ...\right)}   
where the correction term is obtained from the last two terms in (\ref{1li}) with two derivatives acting on the background\footnote{More generally, the leading corrections coming from the 2n$^{th}$ derivative term in the effective action to the graviton propagator denoted by ellipses are proportional to $c^n H_*^{2n}/\mplf^{2n}$.}. This expansion breaks down precisely when 
\eq{}{H_* \sim \frac{\mplf}{\sqrt N}} 
implying the bound (\ref{hbound}) that the scale of inflation cannot be greater than the strong coupling scale $M_{**}$. Furthermore, as seen in the previous section, lower spin species that couple to the trace of the energy momentum tensor do indeed affect the spectrum of tensor and curvature perturbations in such a way that the usual single field tensor to scalar consistency relation is violated (\ref{csv}) due to the process dependence of $M_*$.

\section{Concluding remarks}

In this note, we have elaborated upon various consequences of the fact that characteristic strength of gravity at a given energy ($M_*$) and its strong coupling scale ($M_{**}$) are in general different from each other and the macroscopically determined $\mplf$, particularly as they relate to inferring absolute scales from cosmological observations. This is because universally coupled species (defined as all particles that can mediate \textit{tree level} interactions between conserved sources) affect the strength of gravity at distances smaller than their Compton wavelength. Moreover, all species present (universally coupled, hidden or otherwise) drag down the strong coupling scale -- where a classical description of geometry breaks down -- as $M_{**} = \mplf/N$, where $N$ is the total number of species. This necessarily bounds the scale of inflation from above. As stressed, although observables are dimensionless ratios of quantities measured at a fixed scale and thus independent of the units in which they are expressed, inferring an absolute scale for inflation from any detection of primordial tensors is complicated by the fact that we simply do not know $M_*$ and $M_{**}$ during inflation. Along the way, we made an observation of possible wider interest, namely that KK gravitons necessarily evade the Higuchi bound on any consistent compactification of higher dimensional Einstein gravity -- a result guaranteed by the healthiness of the Einstein-Hilbert action in any number of dimensions. We understand why this is so in the appendices.

\acknowledgements \noindent SP is supported by funds from the Swiss National Science Foundation.

\appendix

\section{de Sitter branes in an empty bulk}
Consider Einstein gravity in 5D. The most general metric ansatz that preserves 4D homogeneity and isotropy is given by
\eq{met}{g_{AB} = -n^2(t,y)dt^2 + a^2(t,y)\delta_{ij}dx^idx^j + b^2(t,y)dy^2}
where upper case Latin indices run from 0 to 4. Coupled to conserved sources, the Einstein equations are given by
\eq{}{G_{AB} = \kappa_5^2\,T_{AB}}
where $\kappa_5^2$ relates to the 5-dimensional Planck mass as $\kappa_5^2 = {1}/{\mpl^3}$. Assuming a constant sized extra dimension, the ansatz (\ref{met}) can be further factorized as
\begin{eqnarray*}
n(t,y) = \w n(y),~~~~~ a(t,y) = a_0(t)\w a(y),~~~~~ b(t,y) = \w b(y).
\end{eqnarray*}
If the EM tensor $T^A_B$ has a vanishing 05 component, the corresponding component of the Einstein tensor $G^A_B$
\begin{eqnarray}
\label{05}
G^0_5 &=& -\frac{3}{n^2}\left[\fp n \ft a - \frac{\dot a'}{a} \right]
\end{eqnarray}
implies the constraint equation  $\w n(y) \propto \w a(y)$ which we can further normalize as $\w n \equiv \w a$. This factorizability further allows us to make the gauge choice $\w b \equiv 1$, where the physical size of the extra dimension now is encoded in the range of the coordinate interval $y$. Finally, making the further ansatz of an induced dS like solution $a_0 = e^{H t}$, the line element becomes
\eq{ans}{ds^2 = \w a(y)^2\left(-dt^2 + e^{Ht}dx^idx^i\right) + dy^2.}
The remaining Einstein equations are
\begin{eqnarray}
\label{00}
\frac{H^2}{\w a^2} &=& \fpp{\w a} + \frac{\w a'^2}{\w a^2} - \frac{\kappa_5^2}{3}T^0_0\\ \label{ii} \frac{H^2}{\w a^2} &=& \fpp{\w a} + \frac{\w a'^2}{\w a^2} - \frac{\kappa_5^2}{3}T^i_i \\ \frac{H^2}{\w a^2} &=& \frac{\w a'^2}{\w a^2} - \frac{\kappa_5^2}{6} T^5_5
\end{eqnarray}
where we have decomposed the EM tensor as a bulk contribution and brane contributions:
\eq{}{T^A_C = T^A_{(b)\,C} + \sum_i T^A_{(i)\,C}\,\delta(y-y_i),} 
\eq{emt}{T^A_{(b)\,B} = {\rm diag}\left(-\rho_{b},p_b,p_b,p_b,r_b\right)}
\eq{bemt}{T^A_{(i)\,B} = \delta(y - y_i)\,{\rm diag}\left(-\rho_i,p_i,p_i,p_i,0\right),}
where we recall that the Jacobian factor that ordinarily appears dividing the delta function in (\ref{bemt}) is $b \equiv 1$. We focus on solutions on an orbifold topology $\mathbb R^4 \times S^1/\mathbb Z_2$. Since the metric has to be continuous everywhere although its derivatives can jump, we find from (\ref{00}) and (\ref{ii}) the junction conditions at the branes \cite{Binetruy, ACK}:
\eq{jc}{ [\w a]_{y_i} = -\frac{\rho_i}{3}\w a(y_i) = \frac{p_i}{3}\w a(y_i) }
where $[\w a]_{y_i} := \w a(y_i)_+ - \w a(y_i)_-$. Therefore the ansatz (\ref{ans}) requires that the EM tensors on the branes be cosmological constant (CC) like. For an empty bulk, one obtains the following warped solution for $\mathbb R^4 \times S^1/\mathbb Z_2$:
\eq{5ddec}{ds^2 = \frac{\left(1 + H|y|\right)^2}{H^2\tau^2}\left( -d\tau^2 + dx_1^2 + dx_2^2 + dx_3^2\right) + dy^2 } 
i.e. $\w a(y) = 1 + H|y|$, where we have immediately switched to conformal coordinates and where the junction conditions imply 
\eq{hans}{H = -\kappa_5^2\Lambda_0/6}
with $\Lambda_0$ being the tension of the brane at $y=0$, taken to be negative so that one has an expanding induced dS solution. The junction conditions at the second brane at $y = y_c$ implies the relations between the tensions $\Lambda_c = -\Lambda_0/(1 + H|y_c|)$, or 
\eq{brel}{\Lambda_c = -\frac{\Lambda_0}{\w a_c},~~~~ \w a_c = 1 + H|y_c|}
We take $\Lambda_c$ as the tunable parameter of our solution. From (\ref{brel}) and (\ref{hans}), this implies the inter-brane separation
\eq{}{|y_c| = \frac{6(\Lambda_0 + \Lambda_c)}{\kappa_5^2\Lambda_0\Lambda_c}.}
Thus we see that in the limit $\Lambda_c \to 0$, one can make the size of the extra dimension as large as one wants whilst keeping $H^2 = \Lambda_0^2\kappa_5^4/36$ fixed so that our solution consistently attains:
\eq{Hhier}{H^2 \gg \frac{1}{y_c^2}} 
Therefore, an observer localized on one of the branes will observe an induced dS metric with propagating KK gravitons with masses that can be made much lighter than $H^2$. We infer the latter from studying the equation of motion for a minimally coupled scalar $\square_5\phi = 0$. With the metric (\ref{5ddec}), this becomes
\eq{}{\square_4\phi + \partial_5^2\phi + [2\theta(y) - 1]\frac{4H}{1 + H|y|}\partial_5\phi =0.}
Factorizing the mode function solutions as $\phi(\tau,x,y) = \phi_4(\tau,x)\w\phi(y)$, the KK spectrum can be read off as the allowed eigenvalues of the equation
\eq{}{\partial_5^2\w \phi + [2\theta(y) - 1]\frac{4H}{1 + H|y|}\partial_5\w\phi = -m^2\w\phi}
which has two independent solutions
\eq{}{\w a^{-3/2} J_{3/2}\left[m/H + m|y| \right],~~~ \w a^{-3/2} Y_{3/2}\left[m/H + m|y| \right] }
where the boundary conditions on the solutions implies a spectrum of states with mass eigenstates separated by the zeros of the Bessel functions \cite{Randall}, i.e. such that up to factors of order unity for small $n$ (or for large $n$):
\eq{}{m^2 \sim \pi^2\frac{n^2}{y_c^2}}
so that given (\ref{Hhier}) one can always find KK gravitons that explicitly evade Higuchi's bound in this construction.

\section{How KK gravitons evade the Higuchi bound}

Consider small perturbations around the background metric (\ref{5ddec}):
\eq{}{g_{AB} = g^{(0)}_{AB} + h_{AB}}
The perturbed Einstein equations in 5D are given by
\begin{eqnarray}
\nn
&&-\frac{1}{2}\square h_{AB} -\frac{1}{2}\nabla_A\nabla_B h + \frac{1}{2}\nabla_C\nabla_A h^C_B + \frac{1}{2}\nabla_C\nabla_B h^C_A \\ \label{mgf} && + \frac{g^{(0)}_{AB}}{2}h^{CD}R_{CD} +\frac{g^{(0)}_{AB}}{2}\square h - \frac{g^{(0)}_{AB}}{2}\nabla_C\nabla_D h^{CD} - \frac{1}{2}R h_{AB} = \kappa_5^2 \delta T_{AB}
\end{eqnarray}
In what follows, we will need the non-vanishing components of the Ricci and Riemann tensors (greek indices run from 0 to 3):
\eq{}{R_{\mu\nu} = -2 H g^{(0)}_{\mu\nu}\left(\delta(y) - \frac{\delta(y-y_c)}{\w a_c} \right)}
\eq{}{R_{55} = -8 H \left(\delta(y) - \frac{\delta(y-y_c)}{\w a_c} \right)}
\eq{}{R^\mu_{~5\,\nu\, 5} = -2H\delta^\mu_\nu\left(\delta(y) - \frac{\delta(y-y_c)}{\w a_c} \right)}
\eq{}{R^5_{\mu\,\nu\, 5} = 2H g^{(0)}_{\mu\nu}\left(\delta(y) - \frac{\delta(y-y_c)}{\w a_c} \right)}
That is, the metric (\ref{5ddec}) is flat everywhere except at the branes. Furthermore, we realize that we can exploit 5D diffeomorphism invariance to work in the gauge 
\eq{}{h_{55} = h_{\mu 5} = 0;~ \nabla_A h^A_B = \frac{1}{2}\nabla_B h}
which corresponds to the usual 5D de Donder gauge, further exploiting the residual gauge symmetry to fix $h_{55}$ and $h_{5\mu}$. We find that with this gauge choice, the $\mu\,\nu$ component of (\ref{mgf}) implies
\begin{eqnarray}
\nn
-\frac{1}{2}\square h_{\mu\nu} + \frac{g_{\mu\nu}^{(0)}}{4} \square h &+& 6H \left(\delta(y) - \frac{\delta(y-y_c)}{\w a_c} \right)h_{\mu\nu} - H g^{(0)}_{\mu\nu}\left(\delta(y) - \frac{\delta(y-y_c)}{\w a_c} \right)h\\ &=& -\kappa_5^2\,h_{\mu\nu}\left(\Lambda_0\delta(y) + \Lambda_c\delta(y-y_c) \right) \end{eqnarray}
However, given the boundary conditions on the brane tensions $\Lambda_c = -\Lambda_0/\w a_c$  (\ref{brel}), we see that the relation (\ref{hans}) enforces a cancellation for the mass terms for $h_{\mu\nu}$ sourced by the background, leaving us with
\eq{}{-\frac{1}{2}\square h_{\mu\nu} + \frac{g_{\mu\nu}^{(0)}}{4} \square h - H g^{(0)}_{\mu\nu}\left(\delta(y) - \frac{\delta(y-y_c)}{\w a_c} \right)h = 0.}
Furthermore, the $5 5$ component of (\ref{mgf}) can be shown to imply 
\eq{}{\frac{1}{4}\square h  - H\left(\delta(y) - \frac{\delta(y-y_c)}{\w a_c} \right)h = 0}
Leaving us with the equation of motion that is absent from any explicit source terms from the background geometry:
\eq{}{\square h_{\mu\nu} = 0}
Expanding $h_{\mu\nu}$ in terms of KK eigenmodes $h^{(n)}_{\mu\nu}$ and integrating over the extra dimension \cite{Giudice}, we will reproduce the equations of motion for each spin-2 KK resonance
\eq{}{\left(\square_4 - M^2_n\right)h^{(n)}_{\mu\nu}=0}
where $\square_4$ is the d'Alembertian on a 4D dS background and where $M_n^2 \propto n^2/y_c^2$ is the corresponding eigenvalue. Since it was explicit background sources that resulted in $h_{00}$ obtaining negative norm states for a certain mass range in \cite{Higuchi}, we see that such terms simply do not arise in our set-up. This is a direct consequence of the fact that the compactification spontaneously breaks 5D Lorentz invariance, resulting in extra sources that cancel contributions from the background geometry that would have otherwise been problematic. Hence there are no propagating ghosts and Higuchi's bound is evaded, as should have been obvious from the outset due to the intrinsic healthiness of higher dimensional Einstein gravity. The general lesson that this exercise is illustrative of is that there is no one generally applicable Higuchi bound, merely a given set of stability conditions to be evaluated on a case by case (theory by theory) basis, and that for some theories, these conditions are trivially satisfied as is evidently the case for the manifestly healthy higher dimensional Einstein gravity.


\begin{thebibliography}{99}

\bibitem{Donoghue} 
  J.~F.~Donoghue,
  ``General relativity as an effective field theory: The leading quantum corrections,''
  Phys.\ Rev.\ D {\bf 50}, 3874 (1994)
  [gr-qc/9405057].

\bibitem{Burgess} 
  C.~P.~Burgess,
  ``Quantum gravity in everyday life: General relativity as an effective field theory,''
  Living Rev.\ Rel.\  {\bf 7}, 5 (2004)
  [gr-qc/0311082].

\bibitem{Antoniadis:2014xva} 
  I.~Antoniadis and S.~P.~Patil,
  ``The Effective Planck Mass and the Scale of Inflation,''
  Eur.\ Phys.\ J.\ C {\bf 75}, 182 (2015)
  [arXiv:1410.8845 [hep-th]].

\bibitem{Dvali3} 
  G.~Dvali,
  ``Black Holes and Large N Species Solution to the Hierarchy Problem,''
  Fortsch.\ Phys.\  {\bf 58}, 528 (2010)
  [arXiv:0706.2050 [hep-th]].

\bibitem{Dvali2} 
  G.~Dvali and M.~Redi,
  ``Black Hole Bound on the Number of Species and Quantum Gravity at LHC,''
  Phys.\ Rev.\ D {\bf 77}, 045027 (2008)
  [arXiv:0710.4344 [hep-th]].

\bibitem{Dvali1} 
  G.~R.~Dvali, G.~Gabadadze, M.~Kolanovic and F.~Nitti,
  ``Scales of gravity,''
  Phys.\ Rev.\ D {\bf 65}, 024031 (2002)
  [hep-th/0106058].

\bibitem{Cheung:2007st} 
  C.~Cheung, P.~Creminelli, A.~L.~Fitzpatrick, J.~Kaplan and L.~Senatore,
  ``The Effective Field Theory of Inflation,''
  JHEP {\bf 0803}, 014 (2008)
  [arXiv:0709.0293 [hep-th]].

\bibitem{Chluba} 
  J.~Chluba, J.~Hamann and S.~P.~Patil,
  ``Features and New Physical Scales in Primordial Observables: Theory and Observation,''
  Int.\ J.\ Mod.\ Phys.\ D {\bf 24}, no. 10, 1530023 (2015)
  [arXiv:1505.01834 [astro-ph.CO]].

\bibitem{Kleban} 
  M.~Kleban, M.~Mirbabayi and M.~Porrati,
  ``Effective Planck Mass and the Scale of Inflation,''
  arXiv:1508.01527 [hep-th].

\bibitem{Higuchi} 
  A.~Higuchi,
  ``Forbidden Mass Range for Spin-2 Field Theory in De Sitter Space-time,''
  Nucl.\ Phys.\ B {\bf 282}, 397 (1987).
  
\bibitem{polnarev}
  A.~G.~Polnarev, ``Polarization and Anisotropy Induced in the Microwave Background by Cosmological Gravitational Waves,''
  Sov. Astron. \textbf{29}, 607 (1985).

\bibitem{Birrell} 
  N.~D.~Birrell and P.~C.~W.~Davies,
  ``Quantum Fields in Curved Space,''
  
\bibitem{Vassilevich:2003xt} 
  D.~V.~Vassilevich,
  ``Heat kernel expansion: User's manual,''
  Phys.\ Rept.\  {\bf 388}, 279 (2003)
  [hep-th/0306138].

\bibitem{Adelberger:2003zx} 
  E.~G.~Adelberger, B.~R.~Heckel and A.~E.~Nelson,
  ``Tests of the gravitational inverse square law,''
  Ann.\ Rev.\ Nucl.\ Part.\ Sci.\  {\bf 53}, 77 (2003)
  [hep-ph/0307284].

\bibitem{Kapner:2006si} 
  D.~J.~Kapner, T.~S.~Cook, E.~G.~Adelberger, J.~H.~Gundlach, B.~R.~Heckel, C.~D.~Hoyle and H.~E.~Swanson,
  ``Tests of the gravitational inverse-square law below the dark-energy length scale,''
  Phys.\ Rev.\ Lett.\  {\bf 98}, 021101 (2007)
  [hep-ph/0611184].

\bibitem{Gasperini} 
  M.~Gasperini,
  ``Cosmology and short-distance gravity,''
  arXiv:1508.06100 [gr-qc].

\bibitem{Giudice} 
  G.~F.~Giudice, R.~Rattazzi and J.~D.~Wells,
  ``Quantum gravity and extra dimensions at high-energy colliders,''
  Nucl.\ Phys.\ B {\bf 544}, 3 (1999)
  [hep-ph/9811291].

\bibitem{Anber}
  M.~M.~Anber, J.~F.~Donoghue and M.~El-Houssieny,
  ``Running couplings and operator mixing in the gravitational corrections to coupling constants,''
  Phys.\ Rev.\ D {\bf 83} (2011) 124003
  [arXiv:1011.3229 [hep-th]];
  M.~M.~Anber and J.~F.~Donoghue,
  ``On the running of the gravitational constant,''
  Phys.\ Rev.\ D {\bf 85}, 104016 (2012)
  [arXiv:1111.2875 [hep-th]].

\bibitem{NBB}
  N.~E.~J.~Bjerrum-Bohr, J.~F.~Donoghue, B.~K.~El-Menoufi, B.~R.~Holstein, L.~Planté and P.~Vanhove,
  ``The Equivalence Principle in a Quantum World,''
  arXiv:1505.04974 [hep-th].

\bibitem{Ozsoy:2014sba} 
  O.~\"Ozsoy, K.~Sinha and S.~Watson,
  ``How Well Can We Really Determine the Scale of Inflation?,''
  Phys.\ Rev.\ D {\bf 91}, no. 10, 103509 (2015)
  [arXiv:1410.0016 [hep-th]].

\bibitem{Namba:2015gja} 
  R.~Namba, M.~Peloso, M.~Shiraishi, L.~Sorbo and C.~Unal,
  ``Scale-dependent gravitational waves from a rolling axion,''
  arXiv:1509.07521 [astro-ph.CO].

\bibitem{Weinberg} 
  S.~Weinberg,
  ``Effective Field Theory for Inflation,''
  Phys.\ Rev.\ D {\bf 77}, 123541 (2008)
  [arXiv:0804.4291 [hep-th]].

\bibitem{Bezrukov} 
  F.~L.~Bezrukov and M.~Shaposhnikov,
  ``The Standard Model Higgs boson as the inflaton,''
  Phys.\ Lett.\ B {\bf 659}, 703 (2008)
  [arXiv:0710.3755 [hep-th]].
 
\bibitem{planckcos} 
  P.~A.~R.~Ade {\it et al.} [Planck Collaboration],
  ``Planck 2015 results. XIII. Cosmological parameters,''
  arXiv:1502.01589 [astro-ph.CO].
 
\bibitem{Binetruy} 
  P.~Binetruy, C.~Deffayet and D.~Langlois,
  ``Nonconventional cosmology from a brane universe,''
  Nucl.\ Phys.\ B {\bf 565}, 269 (2000)
  [hep-th/9905012].

\bibitem{ACK} 
  I.~Antoniadis, S.~Cotsakis and I.~Klaoudatou,
  ``Enveloping branes and brane-world singularities,''
  Eur.\ Phys.\ J.\ C {\bf 74}, no. 12, 3192 (2014)
  [arXiv:1406.0611 [hep-th]].

\bibitem{Randall} 
  L.~Randall and R.~Sundrum,
  ``An Alternative to compactification,''
  Phys.\ Rev.\ Lett.\  {\bf 83}, 4690 (1999)
  [hep-th/9906064].

\end{thebibliography}
\end{document}